# Fully non-scanning three-dimensional imaging using an all-optical Hilbert transform enabled by an optical frequency comb

**Takashi Kato[1, 2], Hirotaka Ishii[1, 2], Kazuhiro Terada[1, 2], Tamaki Morito[1, 2], Kaoru Minoshima[1, 2,*]**
[1.] The University of Electro-Communications (UEC), [2.] JST, ERATO MINOSHIMA Intelligent Optical Synthesizer (IOS).
*Author to whom correspondence should be addressed. Electronic mail: k.minoshima@uec.ac.jp

**Abstract:** This paper demonstrates that the precise phase controllability of an optical frequency comb enables all-optical signal processing for the first time. A novel all-optical Hilbert transform is presented with precise control of relative carrier-phase and envelope of optical pulse train based on frequency control utilizing an optical frequency comb. With the proposed all-optical signal processing method, fully non-scanning one-shot three-dimensional (3D) imaging can be realized with high image resolution. The technique can be applied to coherent phase imaging simultaneously. A precise pair of 90°-phase-shifted optical pulses over the entire spectral bandwidth can be generated based on the precise optical phase controllability of an optical frequency comb, thereby facilitating a real-time and precise all-optical Hilbert transform to obtain amplitude and phase of optical signal in a single shot of ultrashort pulses. In our experiments, we realized single-shot 3D imaging with an uncertainty of 5 μm and obtained a surface profile with a resolution of 200 × 200 pixels.

## 1. Introduction

Three-dimensional (3D) imaging techniques are widely utilized in industrial measurement and biomedical imaging systems. Among such techniques, techniques based on distance imaging methods are attractive because they can provide direct and quantitative results without requiring complex analyses or model assumptions. Furthermore, non-scanning methods are required for various applications because they can be applied to moving targets and can capture ultrafast transient phenomena. However, existing methods are insufficient for performing non-scanning 3D measurements with high accuracy over a large measurement range (i.e., large dynamic range). Previously, we reported a novel method for single-shot non-scanning 3D imaging using optical frequency combs [1]. Based on ultrafast conversions between space, time, and frequency information accomplished using a chirped comb (i.e., coherent chirped ultrashort pulse train generated by an optical frequency comb [2]), non-scanning 3D imaging was achieved. Spectral interferometry can provide a linear time-gating method based on a practical low-power laser, such as a fiber comb. Furthermore, pulse-to-pulse interferometry using optical frequency combs significantly extends the longitudinal measurement range without sacrificing measurement precision based on high-precision pulse-to-pulse separation, which can overcome fundamental measurement tradeoffs to achieve extreme dynamic ranges. Furthermore, the simultaneous imaging of separate surfaces with m-order distances without any loss of precision has been demonstrated for highly reflective targets [1]. However, the previously developed method requires a conventional spectrometer with a 1D diffraction grating to detect interference fringe spectra, meaning line scanning is still required for detecting 2D spectral fringe patterns (i.e., 3D images), even though 3D information can be captured in a single ultrashort pulse. Using a fiber bundle to convert a 2D pattern into a 1D pattern represents a partial solution to this problem, but transverse image resolution is still limited because the maximum number of fibers in a bundle is limited [3].

To obtain 2D spectral images with high spatial resolution, hyperspectral imaging techniques are required. There are three types of existing hyperspectral imaging methods. The first is to slice an image into small regions and diffract each region using a grating without spatial overlap (i.e., slicing images with a mirror or splitting images using a lens array [4]). The second is to apply spectroscopy to complete images, such as Fourier spectroscopy [5] or variable optical bandpass filters [6]. The third is to perform inverse problem solving, such as using a pattern mask [7] or 2D diffraction grating [8]. In our technique, to detect high-spatial-resolution 3D images in a single shot, the amplitude of the spectral interference fringe must be obtained without scanning. Therefore, we require a hyperspectral imaging method that simultaneously provides high spatial resolution, high spectral resolution, and non-scanning measurement. However, no existing methods satisfy these requirements.

In this study, we developed a novel method for hyperspectral imaging based an all-optical signal processing technique utilizing an optical frequency comb. The phase controllability of an optical frequency comb was applied to all-optical signal processing for the first time in this study. The developed method achieves both high spatial and spectral resolutions without any scanning, meaning it can be applied to real-time interference fringe analysis. Using the proposed technique, a fully non-scanning one-shot 3D imaging method with high spatial resolution is demonstrated.

## 2. One-shot 3D imaging method

The main principle underlying our developed 3D imaging method [1] is that the interference of a pulse pair with different chirp characteristics generates 2D fringe spectra with wavelength-variant spacing. In the resulting spectra, the slowest varying fringe (i.e., the fringe pattern with broadest spacing) appears at a certain wavelength position (i.e., characteristic color) representing the spectral component of the chirp pulse, where the probe pulse overlaps the reference pulse in time. Therefore, if an intensity signal that possesses only the characteristic color portion of a spectrum is detected, its central wavelength indicates the delay time between the probe and reference, which is equivalent to longitudinal shape information. By simultaneously detecting 2D isolated spectral intensity profiles using a camera, a 3D image can be captured without analyzing the beam position or delay time (Fig.1 (a)). Such 2D spectral images can be captured using conventional three charged-coupled-device (3CCD) cameras with color filters, which was already demonstrated in our previous work on 3D imaging with an optical Kerr shutter acting as a time gate [2]. However, in the method developed in this study based on spectral interferometry, the interference fringe is superimposed on the detected spectral intensity profile, as shown in the upper-right plot in Fig. 1(a). Therefore, direct spectral images cannot be obtained by simply detecting interference signals. Furthermore, even post-processing cannot provide spectral images because phase information is missing in the absence of delay scanning. Therefore, we must develop a novel principle for real-time signal processing to obtain the amplitude of the fringe spectrum of each pixel using a one-shot image. To this end, we developed a method to obtain the amplitude of a fringe signal using an optical setup implementing an "all-optical Hilbert transform" (i.e., an analogy of the Hilbert transform that is widely used in the field of signal processing to obtain amplitude and phase values). Such signal processing in the optical domain is enabled by the phase controllability of an optical frequency comb. Finally, generated 2D images with spectral-intensity profiles pass through a pair of long- and short-pass filters, which function similarly to a 3CCD camera, but with linearized wavelength dependencies for precise spectroscopic analysis [9]. The ratio of transmittance between the paired filters determines the wavelength of the spectral-intensity signal (i.e., the characteristic color, as shown in Fig.1 (b)).

To obtain the amplitude and phase of a measured interference signal using the all-optical Hilbert transform, we must generate two interference fringe images with a precise 0°/90° phase shift. Such a phase shift can be introduced by a wave plate, but it is difficult to realize precise phase control over a broad spectral region based on the spectral dependence of a wave plate. Additionally, the precise temporal and spatial matching of an envelope and beam between two phase-shifted pulses is very difficult.

Therefore, we developed a novel method to generate a 90°-phase-shifted pulse pair precisely using an optical frequency comb (Fig. 1(c)). If the ratio between the repetition frequency and carrier-envelope-offset frequency (denoted as $f_{rep}$ and $f_{ceo}$, respectively) of the optical frequency comb is set to 4:1, then the carrier envelope phase difference between adjacent pulse-to-pulse units is precisely maintained at 90° over the entire spectrum. Then, by splitting the phase-stabilized optical frequency comb beam and introducing a fixed time delay corresponding to the pulse-to-pulse separation of $1/f_{rep}$, we can obtain a temporally overlapping pair of high-precision 0°/90° phase-shifted pulses. This method allows us to stabilize both the relative carrier phase and envelope timing of the two phase-shifted pulses generated by actively controlling $f_{rep}$ and $f_{ceo}$ simultaneously. Furthermore, a method using frequency control does not affect the wavefront of the pulses, unlike a method using a wave plate and/or mechanical delay scanning. Experimentally, we generated two phase-shifted pulses using a compact multi-pass cavity (MPC) as a delay line. Subsequently, to cancel environmental fluctuations in optical delay, we detected the interference signal between the two phase-shifted pulses using a balanced photodetector on the side and actively stabilized this signal by controlling the frequency of the comb without any mechanical movement. Simultaneously, $f_{ceo}$ was locked at 25% of $f_{rep}$ by a frequency divider. In this manner, by taking full advantage of the frequency and phase controllability of an optical frequency comb without any mechanical movement, the optical path length and optical phase were stabilized.

## 2. Experimental methods and results

Figure 2(a) presents the experimental setup for 3D imaging using the proposed method. The light source is a lab-made fully phase-stabilized mode-locked Er fiber comb with an Er-doped fiber amplifier (repetition rate ($f_{rep}$): 51 MHz, carrier envelope offset frequency ($f_{ceo}$): 12.7 MHz, pulse width: 65 fs, center wavelength: 1.56 μm, spectral FWHM: 112 nm). This optical setup is based on a Mach-Zehnder interferometer. In this setup, the output beam is split into reference and probe paths. The probe and reference pulses are chirped positively and negatively by inserting a 3 m dispersion-compensating fiber and 3 m single-mode fiber, respectively. After propagating through these long fibers, the probe and reference pulses are extended in time to approximately 16 and 9 ps, respectively, with long tails and the output powers are 84 and 9.8 pJ, respectively. The probe beam is directed to a target and then split into two beams for interference by the pair of 0°/90°-phase-shifted reference pulses. The phase-shifted reference pulses are generated by an MPC composed of two curved mirrors. One of the split reference pulses is directed to the MPC and a time delay equal to $1 / f_{rep}$ is added. The delayed path is then combined with the other path while maintaining perpendicular polarizations. Next, the timing overlap between the two pulse trains is ensured via feedback control of $f_{rep}$ against environmental fluctuations in the optical path length of the MPC. Specifically, $f_{ceo}$ is frequency stabilized to 25% of $f_{rep}$. By using these stabilization processes, we obtain precisely controlled phase-shifted reference pulses. Then, the two interference images generated between the probe and phase-shifted references are split again and introduced into a pair of filters (F1 and F2 in Fig. 2(a)). Finally, the four generated beams are slightly shifted in the transverse direction and four fringe images are simultaneously detected by two InGaAs cameras. Based on simple calculations using these four images (i.e., sum of square and division), a 3D image is obtained in one shot.

To evaluate the performance of the proposed system, we first measured the 3D profile of a flat mirror. Based on the developed all-optical Hilbert transform system, we simply calculated the sum of squares of the pairs of captured phase-shifted images to derive intensity images for both F1 and F2. The intensity ratio between the F1 and F2 images was calculated from these intensity images (Fig. 2(b-1), top). The obtained intensity ratio between F1 and F2 represents spectral information. Therefore, a 2D spectral image can be obtained. To calibrate the spectral information obtained from the intensity ratio, we applied a variable delay to the reference path and performed a series of measurements accordingly (Fig. 2(b-1), bottom). As a result, we obtained the relationship between the intensity ratio and delay (i.e., depth information), which yields a calibration curve (Fig. 2(b-2), top). After this calibration curve is obtained for a given measurement system, delay scanning is no longer necessary for actual measurements. Based on this calibration curve, we can retrieve depth information from the captured interference images of measurement targets at a fixed delay. For smoothing purposes, an 11th-order polynomial approximation was performed to obtain a fitted calibration curve in the target range (red curve in Fig. 2(b-2)). As a result, the standard deviation of the fitted residual was determined to be 4.9 μm within a depth range of 0.6 mm (9.1 μm within a 2.4 mm range). Residual error in the measurement was mainly caused by mismatches between the four interference images (i.e., misalignments between the infrared cameras) and can be addressed through improvements to the optical system. Here, the standard deviation between two phase-shifted pulses without frequency control using an optical frequency comb or the proposed stabilization system was 226 μm. Therefore, the developed system reduced the standard deviation to approximately 50 times.

Finally, we applied the proposed technique to a high-resolution and high-precision non-scanning 3D image with a stabilization system. For a proof-of-principle measurement, we measured the shape of an electrical circuit. Figure 2(c) presents the 3D profile of the sample. Here, the longitudinal length is calculated based on the data obtained from the flat mirror mentioned above and is as a calibration curve (Fig. 2(b-2)). Therefore, the 3D image can be obtained without any scanning. The image resolution and size were determined using an image sensor with a resolution of 256 × 150 pixels. The 3D image was successfully recorded in a single shot of an ultrashort pulse with high image resolution, demonstrating the effectiveness of the proposed system.

## 3. Conclusions

In this paper, we demonstrated that the precise phase controllability of an optical frequency comb enables all-optical signal processing for the first time. An all-optical Hilbert transform was applied in a single shot of ultrashort pulses using a precise pair of phase-shifted pulses over the entire spectral bandwidth generated by an optical frequency comb. With this technique amplitude and phase of optical wave signal can be obtained real-time. Furthermore, through frequency stabilization of the frequency comb parameters, the optical path length can be stabilized without mechanical feedback, meaning the relative envelope and phase between the phase-shifted pulse pair can be stabilized. Using this technique, we can achieve fully non-scanning high-resolution 3D imaging. Note that coherent phase imaging is simultaneously achieved by the all-optical Hilbert transform. The proposed method achieved an uncertainty of 5 μm for one-shot 3D imaging, which is approximately 1 / 50th of the uncertainty achieved without frequency comb stabilization. The 3D surface profile of an electrical circuit was successfully measured. Current uncertainty is limited by our experimental system, which could be improved by completely matching the foci and transverse positions of the four interference images. Furthermore, by using pulse-to-pulse imaging interferometry, surfaces separated by more than a meter could be measured without the need for scanning and without any loss in accuracy [1]. We can conclude that our novel all-optical Hilbert transform method exploiting the high phase controllability of optical frequency combs can be used for a broad range of applications.

## 4. References

This work was supported by grants from the JST, ERATO MINOSHIMA Intelligent Optical Synthesizer (JPMJER1304) and JSPS KAKENHI (19K15464).

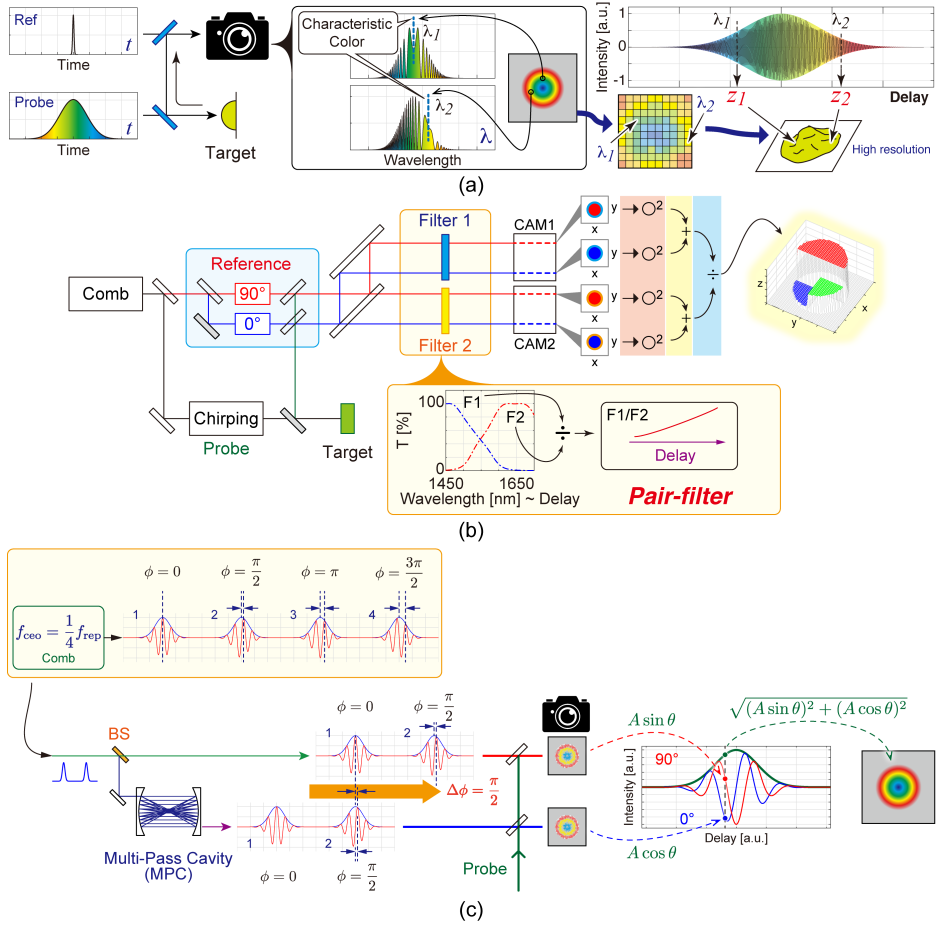

Fig. 1 (a) Principle for 3D imaging with high spatial resolution in a single shot. In the proposed system, we capture interference images between the reference and probe pulses with different chirp characteristics using a camera (a chirped probe and chirp-free reference pulses are shown). Next, we obtain an interference image at each pixel included the spectral dependencies of the fringe pattern, where its characteristic color indicates longitudinal shape information. It should be noted that only the signal near the characteristic color region remains after detection by the detector based on destructive interference between wavelength components caused by high fringe frequencies in regions outside the characteristic color region when spectral interference is integrated. Therefore, the captured image represents a distribution of the interference signals of characteristic color. However, the interference fringe is superimposed on the detected spectral intensity profile, as shown in the upper-right plot in (a). Therefore, a direct spectral image cannot be obtained by simply detecting interference signals. To obtain the amplitude and phase of the fringe signals, signal processing by performing a Hilbert transform in one shot is necessary. Finally, by measuring the characteristic color at each pixel, we can obtain the 3D shape information of the target with high spatial resolution in a single shot. (b) Schematic of the proposed one-shot 3D imaging method. By using a Mach-Zehnder interferometer, we obtain an interference image of the sample. Here, we generate 0°/90°-phase-shifted reference pulses for performing an all-optical Hilbert transform using a phase-controlled optical frequency comb. After interfering the reference and probe pulses, two interference images are obtained. These two interference images are then split again and each pair passes through two color filters (i.e., a pair of short- and long-pass filters, denoted as F1 and F2, respectively). Finally, we capture four interference images using cameras. After calculating simple sums of squares and divisions, we obtain a high-spatial-resolution 3D image in one shot. (c) Principle of the all-optical Hilbert transform using high-precision phase controllability of $f_{\rm rep}$ and $f_{\rm ceo}$. To obtain the amplitude and phase of the characteristic color fringe in real time, we perform a Hilbert transform on the interference images. The amplitude of an interference signal can be calculated from the absolute value of the corresponding Hilbert-transformed signal. The Hilbert-transformed signal provides a complex value whose real and imaginary components are 0°- and 90°-phase-shifted signals, respectively. Through this process, we can obtain the amplitude of the signal in a single shot. We generate 0°- and 90°-phase-shifted pulses using an optical frequency comb, where $f_{\rm rep}$ and $f_{\rm ceo}$ are stabilized to a ratio of 4:1. Then, every pulse-to-pulse carrier phase of the comb precisely changes by 90°. Therefore, by splitting the beam and adding a delay corresponding to the pulse-to-pulse separation using an MPC, we obtain a pair of high-precision 0°/90°-phase-shifted pulses. After interfering the probe and phase-shifted reference pulses, we obtain a pair of 0°/90°-phase-shifted interference images. These signals correspond to a pair of sine and cosine functions. Therefore, by calculating the sum of squares of sine and cosine images, an amplitude image can be obtained in a single shot. Note that coherent phase imaging is achieved simultaneously.

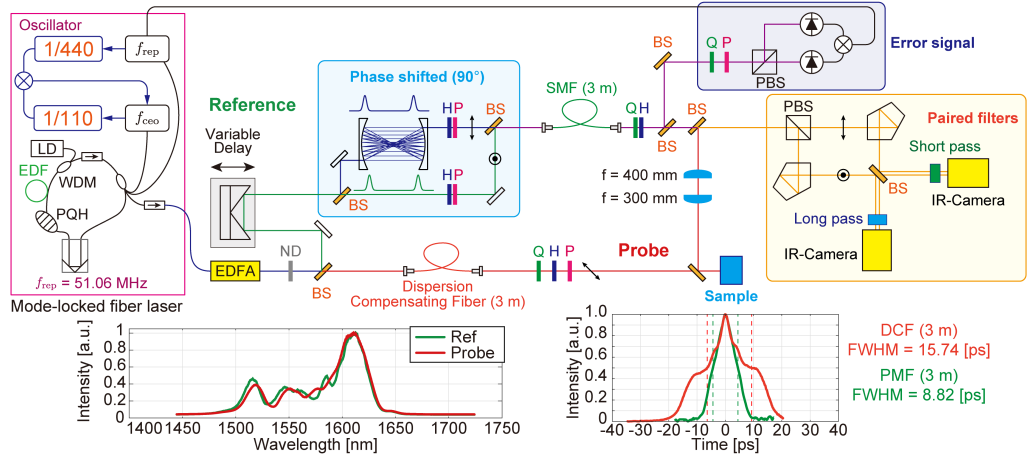

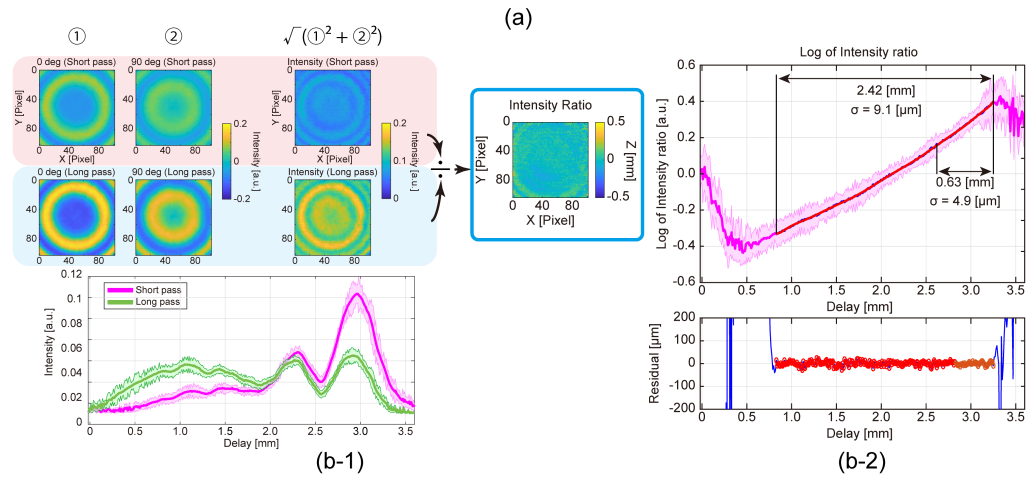

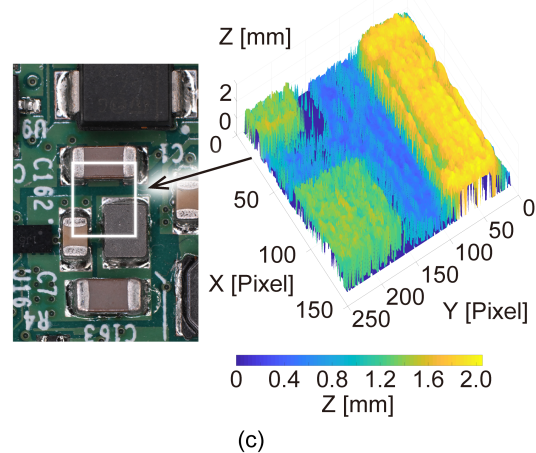

Fig. 2 (a) Experimental setup. The insets show the spectra and pulse temporal shapes of the reference and probe pulses obtained by an optical spectrum analyzer and autocorrelator. (b-1) Upper images show the experimentally obtained raw images (four images), amplitude images (two images), and intensity ratio images, which were obtained using a flat mirror as a target. First, we capture four interference images using two cameras (Fig. 1(b) and Fig. 2(a)). After calculating the sum of squares of the 0°/90°-phase-shifted images for each long and short filter in (a), the amplitude images are obtained. After calculating the ratio of the amplitude images, we obtain an intensity ratio image corresponding to the 3D profile of the target. The lower plot shows the mean value of the amplitude for each image obtained by varying the delay between the probe and reference pulses. (b-2) Upper plot shows the intensity ratio with changing delay, which represents the calibration curve for obtaining the delay (i.e., depth information). Magenta and translucent areas show the mean values and standard deviations, respectively. The red line shows the calibration curve obtained via polynomial approximation. The lower plot shows the residuals between the measured and fitted curves in the upper plot. (c) Resulting 3D profile with an electrical circuit as a measurement target.